\documentclass[aps,twocolumn,superscriptaddress,amsmath,amsfonts,showkeys,
               floatfix]{revtex4}

\usepackage{epsfig,psfrag}
\usepackage{CJK}

\usepackage{graphicx}
\usepackage{dcolumn}
\usepackage{bm}
\usepackage{amssymb}
\usepackage{natbib}

\begin{document}
\title{Infrared optical absorption of Fr$\ddot{o}$hlich polaron in metal halide perovskites}
\author{Yu Cui}
\affiliation{Tianjin Key Laboratory of Low Dimensional Materials Physics and Preparing Technology, Department of Applied Physics, School of Science, Tianjin University, Tianjin 300354 China}
\author{Xiao-Yi Liu}
\affiliation{Tianjin Key Laboratory of Low Dimensional Materials Physics and Preparing Technology, Department of Applied Physics, School of Science, Tianjin University, Tianjin 300354 China}
\author{Xu-Fei Ma}
\affiliation{Tianjin Key Laboratory of Low Dimensional Materials Physics and Preparing Technology, Department of Applied Physics, School of Science, Tianjin University, Tianjin 300354 China}
\author{Jia-Pei Deng}
\affiliation{Tianjin Key Laboratory of Low Dimensional Materials Physics and Preparing Technology, Department of Applied Physics, School of Science, Tianjin University, Tianjin 300354 China}
\author{Yi-Yan Liu}
\affiliation{Tianjin Key Laboratory of Low Dimensional Materials Physics and Preparing Technology, Department of Applied Physics, School of Science, Tianjin University, Tianjin 300354 China}
\author{Zhi-Qing Li}
\affiliation{Tianjin Key Laboratory of Low Dimensional Materials Physics and Preparing Technology, Department of Applied Physics, School of Science, Tianjin University, Tianjin 300354 China}
\author{Zi-Wu Wang*}
\email{wangziwu@tju.edu.cn}
\affiliation{Tianjin Key Laboratory of Low Dimensional Materials Physics and Preparing Technology, Department of Applied Physics, School of Science, Tianjin University, Tianjin 300354 China}

\begin{abstract}
The formation of Fr$\ddot{o}$hlich polaron in metal halide perovskites, arising from the charge carrier-longitudinal optical (LO) phonon coupling, has been proposed to explain their exceptional properties, but the effective identification of polaron in these materials is still a challenge task. Herein, we theoretically present the infrared optical absorption of Fr$\ddot{o}$hlich polaron based on Huang-Rhys model. We find that multiphonon overtones are appeared as the energy of incident photon matches the multiple LO phonons, wherein the average phonon numbers of a polaron can be directly evaluated by the order of the strongest overtone. These multiphonon structures sensitively depend on the scale of electronic distribution in the ground state and the dimensionality of the perovskite materials, which gives the enlightenment for the effective modulation of competing processes between the polaron formation and carrier cooling. Moreover, the order of the strongest overtone shifts to the higher ones with temperature, providing a potential proof of the carriers mobility affected by LO phonons scattering. The present model not only suggests a direct way to verify Fr$\ddot{o}$hlich polaron, but also enriches the understanding of the polaron properties in metal halide perovskites.
\end{abstract}
\maketitle

\section{Introduction}
Fr$\ddot{o}$hlich polaron (or large polaron) formation is originated from the coupling between a macroscopic electronic field induced by the polar optical phonon modes in solid and a charge carrier. So the polaron was also called as the special carrier dressed by the $``phonon$ $cloud"$ (the induced nuclear polarization), reflecting by the average phonon numbers\cite{w1,w2,w3}. This type of polarons has aroused more and more enthusiasm in metal halide perovskites (MHP) because it has been used to unlock the key puzzles regarding the exceptional properties\cite{w4,w6,w7,w8}, such as the moderate carrier mobility, large carrier diffusion length, long carrier lifetime and high defect tolerance. Therefore, the identification on this polaron in MHP becomes one of the crucial problems in both experimental and theoretical studies.

In the past few years, the formation of Fr$\ddot{o}$hlich polaron in perovskite materials has been probed in spectral measurements: following the polaron formation dynamics in time-resolved optical Kerr effect spectroscopy\cite{w9}, ultrafast Terahertz (THz) spectroscopy\cite{w10} as well as time-resolved two-photon photoemission and transient reflectance spectroscopies\cite{w11}; examining Pb-I structural dynamics in time-domain Raman spectroscopy\cite{w13}; studying the electronic structure in angle-resolved photoelectron spectroscopy\cite{w14}. On the other hand, large polarons can be also speculated by other measurements such as the mobility of charge carrier scaling with temperature in the transport\cite{w15} and Hall effect measurements\cite{w16}. In the theoretical aspect, large polarons forming predominantly from the distortion or deformation of the crystal structure with the first-principle calculation have also been widely studied\cite{w5,w17,w18}. Despite these experimental and theoretical evidences have confirmed the formation of large polaron in perovskite materials, the detail knowledge and direct distinction for the ``phonon cloud"  of a polaron, which is the crucial ingredient determining the nature of the polaron, have not been well understood to date.
\begin{figure}
\includegraphics[width=3.5in,keepaspectratio]{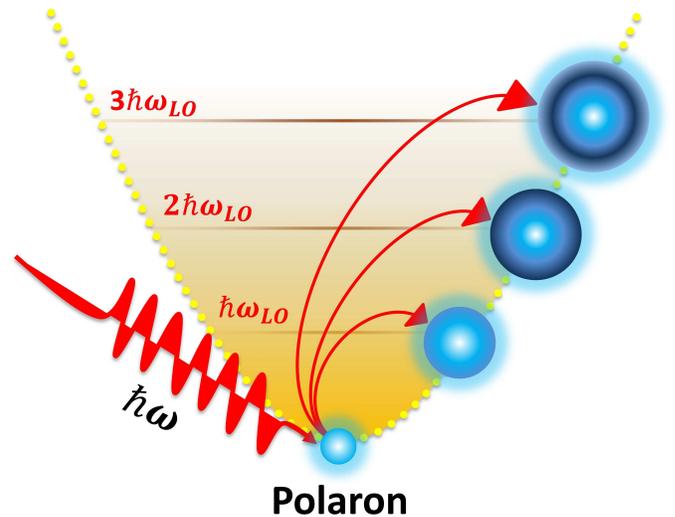}
\caption{\label{compare} The schematic diagram of optical absorption of the polaron, where $\hbar \omega$ is the incident photon energy and  $\hbar {\omega _{LO}}$ is the longitudinal optical (LO) phonon energy.}
\end{figure}

In this paper, we theoretically study the infrared optical absorption of Fr$\ddot{o}$hlich polaron in the three-dimensional (3D) and two-dimensional (2D) MHP materials stemming from the strong electron-longitudinal optical (LO) phonon modes coupling. We find that the ``phonon cloud" of polaron can be directly reflected by multiple LO phonons structure of the optical absorption spectroscopy, in which the average phonon numbers described very well by the dimensionless Huang-Rhys factors are directly linked to the order of the strongest absorption peak. Moreover, the strongest absorption peak shifts to the higher order with decreasing the electronic distribution radius in the ground state, indicating that the range of the electronic distribution plays an important role in determining the electron-LO phonon coupling. Meanwhile, the temperature dependence of the polaron absorption processes is also discussed.
\section{The theoretical model}
\begin{figure}
\includegraphics[width=3.8in,keepaspectratio]{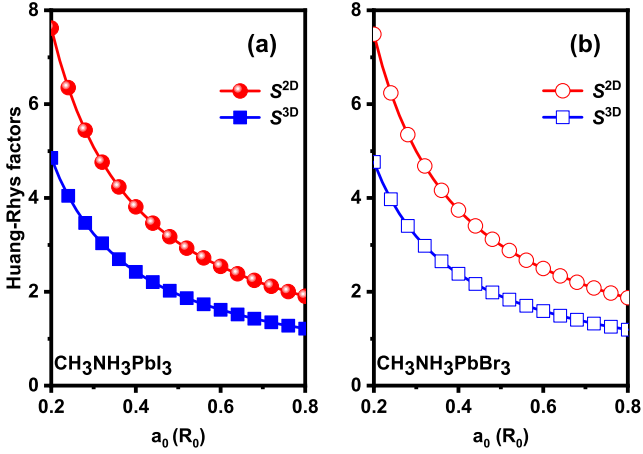}
\caption{\label{compare} Huang-Rhys factors in CH$_3$NH$_3$PbI$_3$ (a) and CH$_3$NH$_3$PbBr$_3$ (b) as a function of the electronic distribution radius in the ground state for different dimensionalities. $\mathcal{S}^{2D}$ and $\mathcal{S}^{3D}$ represent Huang-Rhys factors in two- and three-dimensional perovskite materials, respectively.}
\end{figure}
As schemed in Fig. 1, the absorption of light by polarons can be described as follows: when the incident photon energy $\hbar \omega$ matches the multiple LO phonons energy $p\hbar\omega_{LO}$, the polaron system will undergo transitions from the ground state $|{\Psi _i}\rangle= |{\psi _{1s}}\rangle|{\chi _{i{n'_{\bf{q}}}}}\rangle$ with energy $E_i$ towards a series of excited states $|{\Psi _f}\rangle= |{\psi _{1s}}\rangle|{\chi _{f{n_{\bf{q}}}}}\rangle$ with energy $E_f=E_i+p\hbar\omega_{LO}$. $|\psi _{1s}\rangle$ is the ground state of polaron, and $|{\chi _{i{n'_{\bf{q}}}}}\rangle$ $(|{\chi _{f{n_{\bf{q}}}}}\rangle)$ is the initial (final) lattice wave function, where $n'_{\bf{q}}$ ($n_{\bf{q}}$) is the occupation number for phonons with the wave vector $\bf{q}$.
 Note that the optical absorption discussed here is one polaron behavior rather than many-polaron gas. As is typical of multiphonon transition process, it was excellently studied by the Huang-Rhys model\cite{w24,w25,w26}. By means of this well-known model, the probability of optical absorption for the light with frequency $\omega$ can be expressed as\cite{w20,w21,w22}
\begin{eqnarray}
\mathcal{F}(\hbar \omega ) &=& \sum\limits_f {\frac{{4{\pi ^2}{e^2}\hbar \omega }}{{3{m^2}{\hbar ^2}\varepsilon c}}} {\left| {\left\langle {{\psi _{1s}}} \right|{\bf{r}}\left| {{\psi _{1s}}} \right\rangle } \right|^2}\nonumber\\
&&\times{\left| {\left\langle {{{\chi _{fn_{\bf{q}}}}}}
 \mathrel{\left | {\vphantom {{{\chi _{fn}}} {{\chi _{i{n'_{\bf{q}}}}}}}}
 \right. \kern-\nulldelimiterspace}
 {{{\chi _{i{n'_{\bf{q}}}}}}} \right\rangle } \right|^2}\delta ({E_f} - {E_i} - \hbar \omega ),
\end{eqnarray}
$e$ is the charge carrier, $\hbar\omega$ is the photon energy, $m$ is the effective mass of the polaron, $\varepsilon$ is the permittivity, $c$ is the velocity of light, and $\bf{r}$ is the position variables for polaron.
\begin{figure*}[t]
\centering
\includegraphics[width=7in,keepaspectratio]{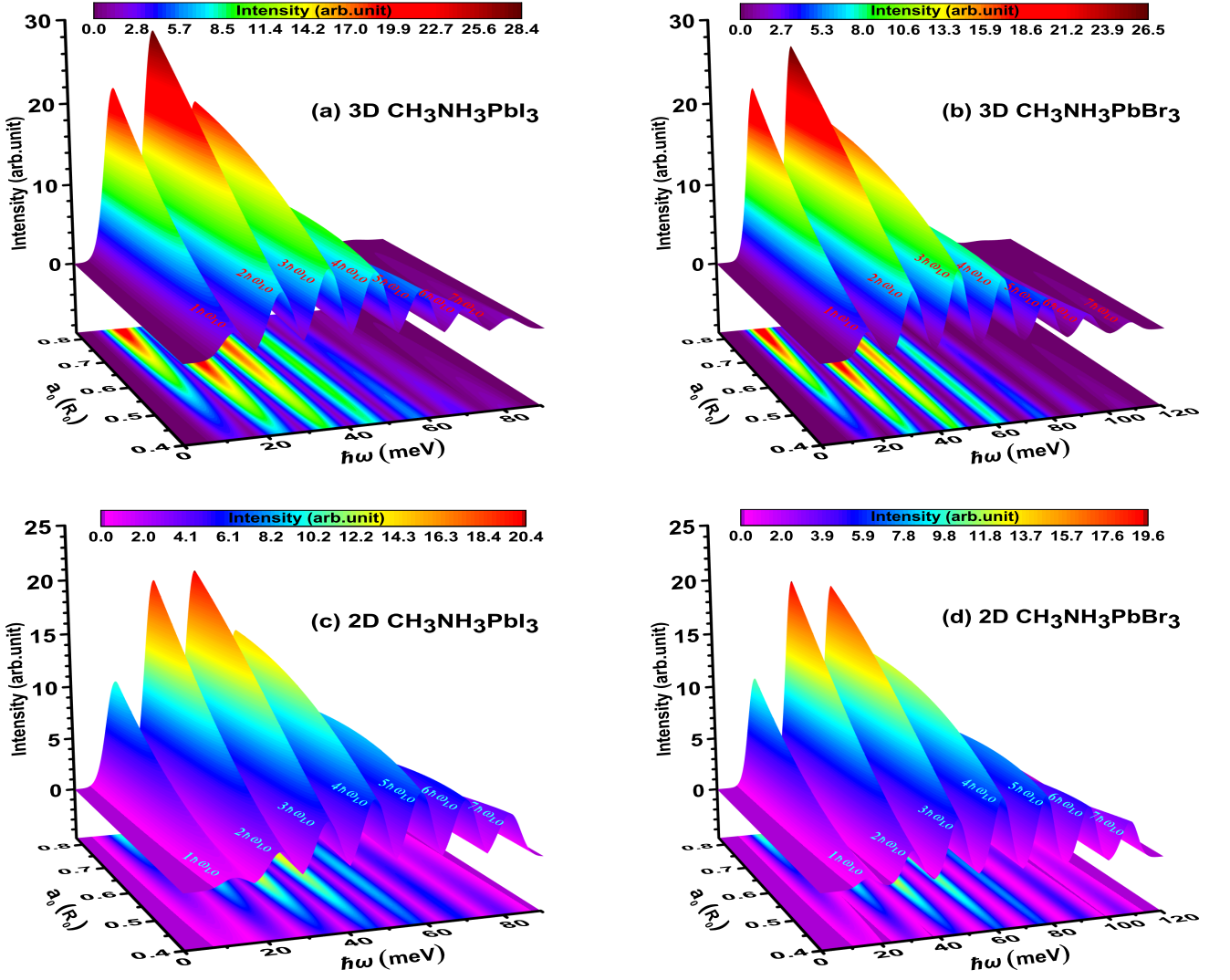}
\caption{Multiple LO phonons infrared optical absorption spectra in 3D CH$_3$NH$_3$PbI$_3$ (a), 3D CH$_3$NH$_3$PbBr$_3$ (b), 2D CH$_3$NH$_3$PbI$_3$ (c) and 2D CH$_3$NH$_3$PbBr$_3$ (d) for different electronic distribution radiuses in the ground state, at $T$ = 77 K.}
\end{figure*}

For the ground state of polaron, a Gaussian wave function is considered\cite{w23}:
\begin{equation}
\psi _{1s}^\lambda (r) = {\mathcal{C}_\lambda }\exp ( - \frac{{{r^2}}}{{4a_0^2}}),
\end{equation}
where the normalization constant ${\mathcal{C}_\lambda } = {[1/(\sqrt {2\pi } {a_0})]^{3/2}}$ for $\lambda$ = $3{\rm D}$ and ${\mathcal{C}_\lambda } = {1/(\sqrt {2\pi } {a_0})}$ for $\lambda$ = $2{\rm D}$ ($\lambda$ = $3{\rm D}$ and $\lambda$ = $2{\rm D}$ represent the 3D and 2D materials, respectively.). $a_0$ represents the radius of electronic distribution in the ground state. Upon substitution of Eq. (2), Eq. (1) will be simplified as
\begin{eqnarray}
{\mathcal{F}^{\lambda}}(\hbar \omega ) &=& \sum\limits_f {{\Re _\lambda }}a_0^2\hbar \omega {\left| {\left\langle {{{\chi _{fn_{\bf{q}}}}}}
 \mathrel{\left | {\vphantom {{{\chi _{fn}}} {{\chi _{i{n'_{\bf{q}}}}}}}}
 \right. \kern-\nulldelimiterspace}
 {{{\chi _{i{n'_{\bf{q}}}}}}} \right\rangle } \right|^2}\nonumber\\
 &&\times \delta ({E_f} - {E_i} - \hbar \omega ).
\end{eqnarray}
 ${\Re _{3D}}=32\pi {e^2} /(3{m^2}{\hbar ^2}\varepsilon c)$ and ${\Re _{2D}}=2\pi^3 {e^2} /(3{m^2}{\hbar ^2}\varepsilon c)$ are the optical constants. The overlap between the initial $|{\chi _{i{n'_{\bf{q}}}}}\rangle$ and final $|{\chi _{f{n_{\bf{q}}}}}\rangle$ lattice vibration produced by the lattice displacement occurring during the transition process is responsible for the multiple phonons structure, which is linked to the electron-phonon coupling.
Substituting the expression of the square of the overlap integral ${\left| {\left\langle {{{\chi _{fn_{\bf{q}}}}}}
 \mathrel{\left | {\vphantom {{{\chi _{fn}}} {{\chi _{i{n'_{\bf{q}}}}}}}}
 \right. \kern-\nulldelimiterspace}
 {{{\chi _{i{n'_{\bf{q}}}}}}} \right\rangle } \right|^2}$ shown in the Appendix A into Eq. (3), we finally convert the optical absorption spectra of polaron involving the participation of $p$ phonons at finite temperature into
\begin{eqnarray}
{\mathcal{F}^{\lambda}}(\hbar \omega ) &=& {\Re _{\lambda}} a_0^2\hbar \omega {(\frac{{\overline n_T  + 1}}{{\overline n_T }})^{p/2}}\exp [ - {\mathcal{S}^{\lambda}}(2\overline n_T  + 1)]\nonumber\\
&&\times{I_p}(2{\mathcal{S}^{\lambda}}\sqrt {\overline n_T (\overline n_T  + 1)} )\exp [ - \frac{{{{(\hbar \omega  - p\hbar {\omega _{LO}})}^2}}}{{2{\sigma ^2}}}].\nonumber\\
\end{eqnarray}
 LO phonon is mainly taken into account because it couples with the charge carriers in Fr$\ddot{o}$hlich mechanism as the main source of lattice distortion for MHP materials in not only theories but also experiments\cite{wsy1,wsy2,wsy3,wb3}. $\overline n_T = 1/({e^{\hbar {\omega _{LO}}/{k_B}T}} - 1)$ is the phonon number occupation function where $k_B$ and $\hbar {\omega _{LO}}$ are the Boltzmann constant and LO phonon energy, respectively; $I_p$ is the $p$th order imaginary argument Bessel function with the phonon number $p$; $\sigma$ represents the line-width of the phonon overtone. The strength of electron-phonon coupling has been characterized using the Huang-Rhys (HR) factor ${\mathcal{S}^{\lambda}}$, which plays a nontrival role in determining the multiphonon processes. In the Fr$\ddot{o}$hlich mechanism\cite{w23,wb1,wb2}, HR factor can be calculated by
\begin{equation}
{\mathcal{S}^{\lambda}} = \sum\limits_{\bf{q}} \ell_\lambda(q) {\left( {\frac{\hbar }{{2{m^*}{\omega _{LO}}}}} \right)^{1/2}}\int {\psi _{1s}^{{\lambda} *}(r){e^{i{\bf{q}} \cdot {\bf{r}}}}} \psi _{1s}^{\lambda}(r) d{\bf{r}}
\end{equation}
$\ell_{3D}(q)=4\pi \alpha /(V{q^2})$ and $\ell_{2D}(q)=2\pi \alpha /(A{q})$. The detailed derivations for HR factors, charge carrier-LO phonon coupling, the corresponding defination and expression of parameters are shown in the Appendix B. In the numerical calculation processes, we choose typical MHP CH$_3$NH$_3$PbI$_3$ (MAPbI$_3$) and CH$_3$NH$_3$PbBr$_3$ (MAPbBr$_3$) as examples to discuss the optical absorption processes of the polaron. Throughout this paper, the line-width $\sigma$ = 3 meV is assumed. The adopted values of other parameters for two types of materials are shown in Table I.

\begin{figure*}[t]
\centering
\includegraphics[width=7in,keepaspectratio]{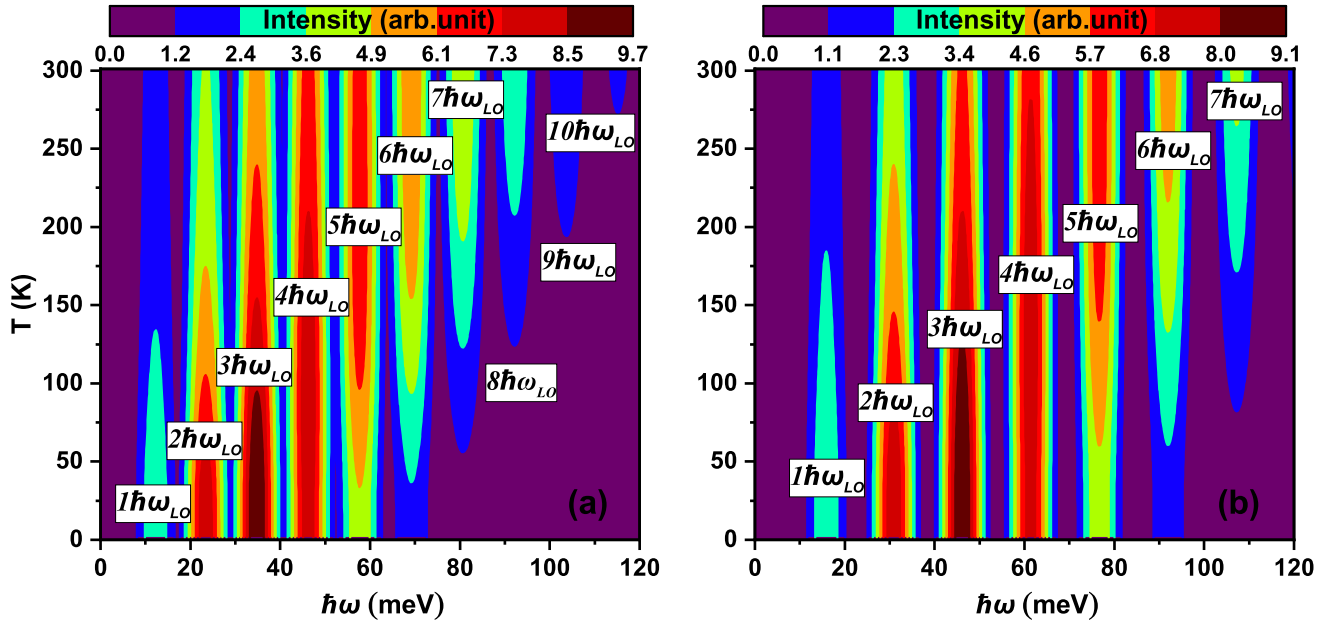}
\caption{The temperature dependences of multiple LO phonons structure for the optical absorption spectra in 3D MAPbI$_3$ (a) and 3D MAPbBr$_3$ (b), respectively, at $a_0=0.4R_0$. }
\end{figure*}

\begin{table}[htbp]
\caption{\label{compare} The adopted parameters for the theoretical calculations. $m_0 = 9.1 \times {10^{-31}}$ kg is the free electron mass. Parameters were taken from Refs. \cite{w27} and \cite{w28}.
	}
\setlength{\tabcolsep}{2mm}{
	\begin{tabular}{ccccccccc}
	
		\hline
		\hline
		Parameter                  & MAPbI$_3$     &  MAPbBr$_3$     \\[0.1ex]\hline
		LO phonon energy ($\hbar\omega_{LO}$) & 11.5 meV\cite{w27} & 15.3 meV\cite{w27} \\
        Coupling constant ($\alpha$) & 1.72\cite{w28} & 1.69\cite{w28} \\
        Effective mass ($m^*$)     &   0.104$m_0$\cite{w28}  &   0.117$m_0$\cite{w28}   \\
		Polaron radius ($R_0$)               & 51 {\AA}\cite{w28} & 43 {\AA}\cite{w28} \\
		\hline
		\hline
	\end{tabular}}
\end{table}

\section{Results and discussion}
HR factor was originally proposed by Huang and Rhys for F-centres in ionic crystals in 1950\cite{w24}, which denotes the average phonon numbers around F-centres, reflecting the magnitude of lattice distortion arising from the strong coupling between electronic states and lattice vibration (phonons). Herein, HR factor is invoked to describe the ``phonon cloud" of the polaron with relaxation energy ${\mathcal{S}^{\lambda}}\hbar\omega_{LO}$. The dependences of HR factors on the electronic distribution radius in the ground state in different dimensionalities are shown in Figs. 2(a) and (b) for MAPbI$_3$ and MAPbBr$_3$, respectively. One can see that HR factors are enlarged from 3D to 2D, which are stemming from that both the reduced dielectric screening and increased degree of quantum confinement are held responsible for the enhancement of electron-phonon coupling. On the other hand, HR factor exhibits a negative correlation with $a_0$. But HR factors vary smoothly as $a_0$ exceeds 0.6$R_0$ ($R_0$ is the polaron radius). This indicates that outside this region the localization feature of electrons has little effect on the electron-phonon coupling. Amounts of researchers have paid attention to the calculation of the formation energy of the large polaron. The obtained values were 12 meV for 3D MAPbI$_3$ and 16$\pm$2 meV for 3D MAPbBr$_3$ in a tight-binding model fitted from the first-principle calculation and temperature-dependent absorption spectroscopy\cite{wz1,wz2}, respectively. These numerical values are in good agreement with the relaxation energy ${\mathcal{S}^{3D}}\hbar\omega_{LO}\approx13.8$ meV for 3D MAPbI$_3$ and ${\mathcal{S}^{3D}}\hbar\omega_{LO}\approx18.2$ meV for 3D MAPbBr$_3$ at $a_0\geq0.6R_0$, implying that the formation energy can be well described by the relaxation energy. As for 2D perovskite species, though there is little direct information about polarons in the experiments, the discoveries of the 2D exciton polarons and self-trapped excitons can reveal the significant polaronic character\cite{w29,w30,w31,w32}. The numerical values of HR factors for 2D perovskites in Fig. 2 are expected to provide insight for the identification on these quasi-particles. In addition, it is shown that HR factors in MAPbBr$_3$ are slightly smaller than that in MAPbI$_3$. It is ascribed to the essence that the distortion of the Pb-Br bond is harder than that of the Pb-I bond determined by the bond length, which is reflected by the coupling parameter $\alpha$ as listed in Table I. We have noticed that the effective phonon modes and relaxation energy could be obtained from the first-principles calculations\cite{w33,w34,w35}, gaining the more accurate evaluations for HR factors. Our results will also provide an important theoretical reference for these calculations.

On the basis of these theoretical values of HR factors, the optical absorption spectra of polaron for MAPbI$_3$ and MAPbBr$_3$ are plotted in Fig. 3 for different electronic distribution radiuses in the ground state. We first give the spectra for MAPbI$_3$ a detailed discussion. In Fig. 3(a), the energy resonance absorption will occur when the energy of the incident photon matches the multiple LO phonons, showing the multiphonon overtones in the infrared (long wavelength) region of spectroscopy, where the strongest overtone corresponds to the most probable state of polaron. And the intensities of overtones follow an asymmetric Gaussian distribution, which are in agreement with the previous measurements in experiments\cite{w36,w10}.
It is worth noting that the order of the strongest overtone $p_{max}$ satisfies the condition of $p_{max}\approx{\mathcal{S}^{3D}}+0.5$, such as $\mathcal{S}^{3D}\approx2.42$ corresponding to $p_{max}=3$ at $a_0=0.4R_0$, but the condition shifts to $p_{max}\approx{\mathcal{S}^{2D}}+1$ for the 2D perovskites, such as $\mathcal{S}^{2D}\approx3.81$ corresponding to $p_{max}=5$ at $a_0=0.4R_0$, shown in Fig. 3(c). This means that the average phonon number of polaron could be estimated directly from the absorption spectra. In terms of this significant outcome, it can be used to explain the overtones along with a linear decay from the first to the higher orders in the multiphonon Roman scattering and THz time-dominant spectroscopy for CsPbBr$_3$ ($\mathcal{S}^{3D}=0.39$)\cite{wb3}. The pronounced redshift behaviors for the position of the spectral weight with the increase of the electron localization are shown, specifically manifested in Fig. 3(a) as (1) the number of the observable overtones reduces from seven at ${a_0} = 0.4{R_0}$, to five at ${a_0} = 0.8{R_0}$; (2) the most probable state transits from the third order at $a_0=0.4R_0$ to the second order at $a_0=0.8R_0$. These phenomena are the direct reflections of the variation of the electron-phonon coupling (HR factors) linked to the charge carrier localization, showing the polaron to be more stable or to be more easily taken off phonon dressing.
The similar trends for MAPbBr$_3$ are illustrated in Figs. 3(b) and (d). Recently, several experiments have proved that the large polaron formation drastically suppresses the hot carriers cooling\cite{ws1,w11,ws3,wq1}. For instance, an envelope of Cs$_4$PbBr$_6$ lattice shell was provided over the CsPbBr$_3$ nanocrystals\cite{wq1}. The enhanced strength of the electron-phonon coupling in such shell material facilitates the formation of polarons with large relaxation energy, and thus depresses the carriers cooling processes. It is, therefore, concluded that if more phonon overtones are appeared in the absorption spectra, the polaron effect will be more pronounced, which could be served as a criterion that how to judge these perovskites materials being suitable for the potential applications in photovoltaic (perfer to the strong polaron effect) or photoelectronic (demanding the fast carrier cooling) devices.

The temperature dependences of optical absorption in the 3D MAPbI$_3$ and 3D MAPbBr$_3$ are shown in Figs. 4(a) and 4(b), respectively. These multiphonon overtones will be divided into two regions: the low orders region including the overtones are in the range of $1\leq{p}\leq\mathcal{S}^{3D}+2$ and the rest of overtones being categorized as the high orders region. In the former region, the intensities of overtones decrease monotonically with temperature. On the contrary, the latter region shows the opposite variational trends and even much higher orders of overtones will be thermally activated. It can be explained by that more thermally activated LO phonons become available with temperature, resulting in (1) the polaron states with more phonons are prone to reach the most probable states, and (2) the orders of overtones increase. The blueshift of the most probable states leads to the shift of the whole absorption spectra towards the higher energy, showing the downside of the relative probability in the low order region and the upside trend for the high order region as given in Figs. 4(a) and (b). More importantly, the polaron, dressed by more thermally activated LO phonons, indicates that the polaron effects are actively strengthened at higher temperature. In the perovskite materials, the long carrier diffusion length does not imply a high charge carrier mobility, yet a detailed understanding of its fundamental mechanisms is still outstanding. According to our theoretical study, the polarons become significantly ``heavier" with temperature, which is not beneficial to transport, especially at room temperature. Thus it provides a potential explanation for the moderate carrier mobility in perovskites, which supports the theoretical predictions about LO phonon mode scattering influencing mobility\cite{w28,wq2}.  Therefore, it may be more appropriate to elucidate the effective mobility by the polaron model in perovskite materials\cite{w28,wz1,w3.9,w4.0,w3.5}. In turn, the measurements of the carrier mobility in experiments could also track the trace of polarons\cite{ww1}. It is additionally revealed that the overtones keep stable outline even at room temperature, implying the polaron states have a very robust lifetime, which will benefit not only the experimental observation but also the potential exploitation of polaron properties in perovskite materials.
Finally, we must emphasize that these perovskite materials undergo two structure phase transitions with temperature, which has an influence on the strength of the electron-phonon coupling and phonon energy\cite{w3.6,w3.7,w3.8}. These effects are not taken into account in the present paper.

\section{Conclusions}
In summary, we theoretically study the infrared optical absorption spectroscopy of Fr$\ddot{o}$hlich polaron in metal halide perovskites. We find that (1) the multiphonon structure can be used as the fingerprint for the polaron, where the average phonon number in the ``phonon cloud" of the polaron is reflected by the order of the strongest overtone in the absorption spectra; (2) the multiphonon structure depends on both the dimensionality of materials and the charge carrier localization sensitively, which give the enlightenment for controlling the competing processes between the polaron formation and carrier cooling; (3) the ``phonon cloud" around the polaron will become larger as the temperature increases and thus affect the carrier mobility, especially at room temperature, which is consistent with the experimental measurements.

\section*{ACKNOWLEDGMENT}
This work was supported by National Natural Science Foundation of China (Grant Nos. 11674241 and 12174283).

\begin{appendix}
\section{The calculation of the overlap integrals}
 Within the adiabatic approximation, the square of the overlap integral of the lattice wave function considering the electron-phonon coupling can be written as\cite{w20,w21,app1}
\begin{eqnarray}
&&{\left| {\left\langle {{{\chi _{fn_{\bf{q}}}}}}
 \mathrel{\left | {\vphantom {{{\chi _{fn_{\bf{q}}}}} {{\chi _{i{n'_{\bf{q}}}}}}}}
 \right. \kern-\nulldelimiterspace}
 {{{\chi _{i{n'_{\bf{q}}}}}}} \right\rangle } \right|^2}\nonumber\\
  &=& {{{\left|\prod\limits_{\bf{q}} {\int {{\chi _{{n_{\bf{q}}}}}(Q)\exp [{\textsl{g}_{\bf{q}}}a_{\bf{q}}^\dag  - \textsl{g}_{\bf{q}}^*{a_{\bf{q}}}]{\chi _{{n'_{\bf{q}}}}}(Q)dQ} } \right|}^2}},\nonumber\\
\end{eqnarray}
$Q$ describes the normal mode of lattice vibration, ${\textsl{g}_{\bf{q}}} = {{{{\cal M}_q}{\rho _{\bf{q}}}} \mathord{\left/{\vphantom {{{{\cal M}_q}{\rho _{\bf{q}}}} {(\hbar {\omega _{LO}}}}} \right.\kern-\nulldelimiterspace} {(\hbar {\omega _{LO}}}})$ is the variational parameter with the coupling element $\mathcal{M}_q$ and the electronic charge density $\rho _{\bf{q}}$. $a_{\bf{q}}^\dag$ ($a_{\bf{q}}$) is the creation (annihilation) operator of a LO phonon with wave vector $\bf{q}$, and ${\chi _{{n_{\bf{q}}}}}(Q)$ (${\chi _{{n'_{\bf{q}}}}}(Q)$) is the free LO phonon states, where $n'_{\bf{q}}$ ($n_{\bf{q}}$) is the occupation number for phonons.
The space translation operator $\exp [ {{\textsl{g}_{\bf{q}}}a_{\bf{q}}^\dag }  - \textsl{g}_{\bf{q}}^*{a_{\bf{q}}}]$ reflects the variation of the equilibrium position on account of the electron-phonon coupling.

The creation and annihilation operators can be expressed as\cite{app2}
\begin{equation}
a_{\bf{q}}^\dag  = \sqrt {\frac{{{\omega _{LO}}}}{{2\hbar }}} (Q - i\frac{P}{{{\omega _{LO}}}}),
\end{equation}
\begin{equation}
{a_{\bf{q}}} = \sqrt {\frac{{{\omega _{LO}}}}{{2\hbar }}} (Q + i\frac{P}{{{\omega _{LO}}}}),
\end{equation}
where $P$ is the momentum operator. Upon substitution of Eqs. (A2) and (A3), Eq. (A1) can be of the form
\begin{eqnarray}
&&{\left| {\left\langle {{{\chi _{fn_{\bf{q}}}}}}
 \mathrel{\left | {\vphantom {{{\chi _{fn_{\bf{q}}}}} {{\chi _{i{n'_{\bf{q}}}}}}}}
 \right. \kern-\nulldelimiterspace}
 {{{\chi _{i{n'_{\bf{q}}}}}}} \right\rangle } \right|^2}\nonumber\\
 &=& {\left|\prod\limits_{\bf{q}} {\int {{\chi _{{n_{\bf{q}}}}}(Q)[\sum\limits_{n = 0}^\infty  {\frac{{{\mathcal{A}^n}}}{{n!}}{Q^n}} {\chi _{n'_{\bf{q}}}}(Q)]dQ} }\right|^2}\nonumber\\
 &=&{\left|\prod\limits_{\bf{q}} {\int {{\chi _{{n_{\bf{q}}}}}[{\chi _{n'_{\bf{q}}}} + \mathcal{A}Q{\chi _{n'_{\bf{q}}}} + \frac{{{\mathcal{A}^2}}}{2}{Q^2}{\chi _{n'_{\bf{q}}}} +  \cdots ]dQ} }\right|^2},\nonumber\\
\end{eqnarray}
where $\mathcal{A} = {[{\omega _{LO}}/(2\hbar )]^{1/2}}({\textsl{g}_{\bf{q}}} - \textsl{g}_{\bf{q}}^*)$. For the change in phonon number between $n_{\bf{q}}$ and $n'_{\bf{q}}$, only transitions in which the phonon number changes at most one contribute to the transition probability. Therefore, only three cases are taken into account for the change of the phonon number:

(1) for the case ${n_{\bf{q}}} = {n'_{\bf{q}}} + 1$, one can get
\begin{eqnarray}
\int {{\chi _{{n'_{\bf{q}}+1}}}[\mathcal{A}Q{\chi _{{n'_{\bf{q}}}}}]dQ}&=&\frac{{\sqrt {{n'_{\bf{q}}} + 1} }}{2} {({\textsl{g}_{\bf{q}}} - \textsl{g}_{\bf{q}}^*)};\nonumber
\end{eqnarray}

(2) for the case ${n_{\bf{q}}} = {n'_{\bf{q}}} - 1$, one can get
\begin{eqnarray}
\int {{\chi _{{n'_{\bf{q}}}-1}}[\mathcal{A}Q{\chi _{{n'_{\bf{q}}}}}]dQ}&=&\frac{{\sqrt {{n'_{\bf{q}}}} }}{2} {({\textsl{g}_{\bf{q}}} - \textsl{g}_{\bf{q}}^*)};\nonumber
\end{eqnarray}

(3) for the case ${n_{\bf{q}}} = {n'_{\bf{q}}}$, one can get
\begin{eqnarray}
&&\int {{\chi _{{n'_{\bf{q}}}}}{\chi _{{n'_{\bf{q}}}}}dQ}  + \frac{{{\mathcal{A}^2}}}{2}\int {{\chi _{{n'_{\bf{q}}}}}{Q^2}{\chi _{{n'_{\bf{q}}}}}dQ}  +  \cdots\nonumber\\
&=& 1 + \frac{{2{n'_{\bf{q}}} + 1}}{8}{({\textsl{g}_{\bf{q}}} - \textsl{g}_{\bf{q}}^*)^2} +  \cdots.\nonumber
\end{eqnarray}

When the transition process involves $p$ phonons, the contribution of these three cases should be made up of $k$ oscillators going down by one quantum and $k+p$ oscillators going up by one quantum. Therefore, the square of the overlap integral can be rewritten as
\begin{widetext}
\begin{eqnarray}
{\left| {\left\langle {{{\chi _{fn_{\bf{q}}}}}}
 \mathrel{\left | {\vphantom {{{\chi _{fn_{\bf{q}}}}} {{\chi _{i{n'_{\bf{q}}}}}}}}
 \right. \kern-\nulldelimiterspace}
 {{{\chi _{i{n'_{\bf{q}}}}}}} \right\rangle } \right|^2}&=&{\left| {\frac{{\sqrt {{n'_{{{\bf{q}}_1}}} + 1} }}{2}({\textsl{g}_{{{\bf{q}}_1}}} - \textsl{g}_{{{\bf{q}}_1}}^*)} \right|^2}{\left| {\frac{{\sqrt {{n'_{{{\bf{q}}_2}}} + 1} }}{2}({\textsl{g}_{{{\bf{q}}_2}}} - \textsl{g}_{{{\bf{q}}_2}}^*)} \right|^2} \cdots {\left| {\frac{{\sqrt {{n'_{{{\bf{q}}_{p + k}}}} + 1} }}{2}({\textsl{g}_{{{\bf{q}}_{p + k}}}} - \textsl{g}_{{{\bf{q}}_{p + k}}}^*)} \right|^2}\nonumber\\
 &&\times{\left| {\frac{{\sqrt {{n'_{{{\bf{q}}_1}}}} }}{2}({\textsl{g}_{{{\bf{q}}_1}}} - \textsl{g}_{{{\bf{q}}_1}}^*)} \right|^2}{\left| {\frac{{\sqrt {{n'_{{{\bf{q}}_2}}}} }}{2}({\textsl{g}_{{{\bf{q}}_2}}} - \textsl{g}_{{{\bf{q}}_2}}^*)} \right|^2} \cdots {\left| {\frac{{\sqrt {{n'_{{{\bf{q}}_{k}}}}} }}{2}({\textsl{g}_{{{\bf{q}}_k}}} - \textsl{g}_{{{\bf{q}}_k}}^*)} \right|^2}\nonumber\\
 &&\times{\left| {\prod\limits_{\bf{q}} {[1 + \frac{{2{n'_{\bf{q}}} + 1}}{8}{{({\textsl{g}_{\bf{q}}} - \textsl{g}_{\bf{q}}^*)}^2} +  \cdots ]} } \right|^2}.
 \end{eqnarray}
 \end{widetext}
For the given $k$ and $p$, there are a large number of distinct transitions, corresponding to different choices of the oscillators. Their total contribution can be obtained by summing Eq. (A5) over all the indices, and dividing afterwards by $k!(k+p)!$. Then Eq. (A5) becomes\cite{w24,w25,w26}
\begin{widetext}
\begin{eqnarray}
{\left| {\left\langle {{{\chi _{fn_{\bf{q}}}}}}
 \mathrel{\left | {\vphantom {{{\chi _{fn_{\bf{q}}}}} {{\chi _{i{n'_{\bf{q}}}}}}}}
 \right. \kern-\nulldelimiterspace}
 {{{\chi _{i{n'_{\bf{q}}}}}}} \right\rangle } \right|^2}&=&\frac{1}{{k!(k + p)!}}{\left\{ {\sum\limits_{\bf{q}} {({n'_{\bf{q}}} + 1)} {\textsl{g}_{\bf{q}}}\textsl{g}_{\bf{q}}^*} \right\}^{p + k}}{\left\{ {\sum\limits_{\bf{q}} {{n'_{\bf{q}}}} {\textsl{g}_{\bf{q}}}\textsl{g}_{\bf{q}}^*} \right\}^k}\exp \left\{ {\sum\limits_{\bf{q}} {[\frac{{2{n'_{\bf{q}}} + 1}}{4}{{({\textsl{g}_{\bf{q}}} - \textsl{g}_{\bf{q}}^*)}^2}]} } \right\}\nonumber\\
 &=&\exp [ - \mathcal{S}(2\overline n_T  + 1)]\sum\limits_k {\frac{{{{[\mathcal{S}(\overline n_T  + 1)]}^{p + k}}{{[\mathcal{S}\overline n_T ]}^k}}}{{k!(k + p)!}}}\nonumber\\
 &=& {(\frac{{\overline n_T  + 1}}{{\overline n_T }})^{p/2}}\exp [ - \mathcal{S}(2\overline n_T  + 1)]{I_p}(2\mathcal{S}\sqrt {\overline n_T (\overline n_T  + 1)} ).
\end{eqnarray}
 \end{widetext}
 $\overline n_T = 1/({e^{\hbar {\omega _{LO}}/{k_B}T}} - 1)$ is the phonon number occupation function where $k_B$ and $\hbar {\omega _{LO}}$ are the Boltzmann constant and LO phonon energy, respectively; $I_p$ is the imaginary argument Bessel function with the phonon number $p$; ${\cal S} = \sum\limits_{\bf{q}} {{{{{\left| {\cal M}_q \right|}^2}{{\left| {{\rho _{\bf{q}}}} \right|}^2}} \mathord{\left/{\vphantom {{{{\left| {{M_q}} \right|}^2}{{\left| {{\rho _{\bf{q}}}} \right|}^2}} {{{(\hbar {\omega _{LO}})}^2}}}} \right.\kern-\nulldelimiterspace} {{{(\hbar {\omega _{LO}})}^2}}}}={\left| {{\textsl{g}_{\bf{q}}}} \right|^2}$ is the Huang-Rhys factor.
\section{The calculation of the Huang-Rhys factor}
The electron-LO phonon coupling in the Fr$\ddot{o}$hlich mechanism can be expressed as\cite{w23,wb1}
\begin{eqnarray}
{\cal H}_{e - p}^\lambda  = \sum\limits_{\bf{q}} {({\cal M}_q^{\lambda *}a_{\bf{q}}^\dag {e^{ - i{\bf{q}} \cdot {\bf{r}}}} + {\cal M}_q^\lambda {a_{\bf{q}}}{e^{i{\bf{q}} \cdot {\bf{r}}}})}
\end{eqnarray}
with
\begin{equation}
{{\left| {{\mathcal{M}^{\lambda}_q}} \right|^2} } = \left\{\begin{array}{rcl}
&{\frac{{{{(\hbar {\omega _{LO}})}^2}}}{{{q^2}}}\frac{{4\pi \alpha }}{V}{(\frac{\hbar }{{2{m^*}{\omega _{LO}}}})^{1/2}}},& (\lambda=3D)\\
&{\frac{{{{(\hbar {\omega _{LO}})}^2}}}{q}\frac{{2\pi \alpha }}{A}{(\frac{\hbar }{{2{m^*}{\omega _{LO}}}})^{1/2}}},& (\lambda=2D)\\
\end{array}\right..
\end{equation}
 $V$ is the crystal volume, $A$ is the crystal area, and $\alpha $ denotes the Fr$\ddot{o}$hlich coupling strength. The source of the lattice polarization is the charge distribution of the polaron described by the Fourier transform
 \begin{equation}
 \rho _{\bf{q}}^\lambda {\rm{  = }}\int {{{\left| {\psi _{1s}^\lambda (r)} \right|}^2}} {e^{i{\bf{q}} \cdot {\bf{r}}}}d{\bf{r}}, (\lambda=3D, 2D).\\
 \end{equation}
Substituting Eq. (B2) as well as (B3) into the expression of Huang-Rhys factor ${\mathcal{S}}$, the Huang-Rhys factor in the Fr$\ddot{o}$hlich mechanism can be calculated by\cite{app1,w24}
\begin{eqnarray}
{\mathcal{S}^{3D}} &=&\sum\limits_{\bf{q}} {\frac{{{{\left| {{\cal M}_q^{3D}} \right|}^2}{{\left| {\rho _{\bf{q}}^{3D}} \right|}^2}}}{{{{\left( {\hbar {\omega _{LO}}} \right)}^2}}}}\nonumber\\
&=& \sum\limits_{\bf{q}} \frac{{4\pi \alpha }}{{V{q^2}}} {\left( {\frac{\hbar }{{2{m^*}{\omega _{LO}}}}} \right)^{1/2}}\int {\psi _{1s}^{{3D} *}(r){e^{i{\bf{q}} \cdot {\bf{r}}}}} \psi _{1s}^{3D}(r) d{\bf{r}}\nonumber\\
&=& \frac{{\alpha {R_0}}}{{\sqrt \pi  {a_0}}}
\end{eqnarray}
for the 3D materials and
\begin{eqnarray}
{\mathcal{S}^{2D}} &=&\sum\limits_{\bf{q}} {\frac{{{{\left| {{\cal M}_q^{2D}} \right|}^2}{{\left| {\rho _{\bf{q}}^{2D}} \right|}^2}}}{{{{\left( {\hbar {\omega _{LO}}} \right)}^2}}}}\nonumber\\
&=& \sum\limits_{\bf{q}} \frac{{2\pi \alpha }}{{A{q}}} {\left( {\frac{\hbar }{{2{m^*}{\omega _{LO}}}}} \right)^{1/2}}\int {\psi _{1s}^{{2D} *}(r){e^{i{\bf{q}} \cdot {\bf{r}}}}} \psi _{1s}^{2D}(r) d{\bf{r}}\nonumber\\
&=& \frac{{\sqrt \pi  \alpha {R_0}}}{{2{a_0}}}
\end{eqnarray}
for the 2D materials. $a_0$ is the electronic distribution radius in the ground state, and ${R_0} = {[\hbar /(2{m^*}{\omega _{LO}})]^{1/2}}$ is the polaron radius.
\end{appendix}

\end{document}